# Pareto Optimal Projection Search (POPS): Automated Radiation Therapy Treatment Planning by Direct Search of the Pareto Surface

Charles Huang, Yong Yang, Neil Panjwani, Stephen Boyd, and Lei Xing

*Abstract—Objective:* Radiation therapy treatment planning is a time-consuming, iterative process with potentially high inter-planner variability. Fully automated treatment planning processes could reduce a planner's active treatment planning time and remove inter-planner variability, with the potential to tremendously improve patient turnover and quality of care. In developing fully automated algorithms for treatment planning, we have two main objectives: to produce plans that are 1) Pareto optimal and 2) clinically acceptable. Here, we propose the Pareto optimal projection search (POPS) algorithm, which provides a general framework for directly searching the Pareto front. *Methods:* Our POPS algorithm is a novel automated planning method that combines two main search processes: 1) gradient-free search in the decision variable space and 2) projection of decision variables to the Pareto front using the bisection method. We demonstrate the performance of POPS by comparing with clinical treatment plans. As one possible quantitative measure of treatment plan quality, we construct a clinical acceptability scoring function (SF) modified from the previously developed general evaluation metric (GEM). *Results:* On a dataset of 21 prostate cases collected as part of clinical workflow, our proposed POPS algorithm produces Pareto optimal plans that are clinically acceptable in regards to dose conformity, dose homogeneity, and sparing of organs-at-risk. *Conclusion*: Our proposed POPS algorithm provides a general framework for fully automated treatment planning that achieves clinically acceptable dosimetric quality without requiring active planning from human planners. *Significance:* Our fully automated POPS algorithm addresses many key limitations of other automated planning approaches, and we anticipate that it will substantially improve treatment planning workflow.

## I. Introduction

EXTERNAL beam radiation therapy involves the delivery of ionizing radiation with the intent to treat diseased tissue while minimizing dose to healthy organs [1], [2]. The goal of treatment planning is then to determine optimal beam angles, shapes, intensities, etc. that satisfy this overall clinical objective. Oftentimes, the treatment planning process represents a bottleneck to high quality patient care, due to the time-consuming nature of the iterative planning process and inter-planner variability.

Prior to treatment, medical images are collected for the patient (e.g. CT, MRI, or PET scans), and physicians contour various anatomical structures on the collected images, including the planning target volume (PTV) and surrounding organs-at-risk (OARs) [3], [4]. For the case of intensity modulated radiation therapy (IMRT), planners determine the appropriate plan configuration (i.e. beam type, beam angle arrangement, etc.) and perform inverse planning to achieve a desired dose distribution or to satisfy various objectives and constraints [3], [4]. Many of these same principles are also used in the case of volumetric modulated arc therapy (VMAT) [5] and have been shown to improve delivery efficiency while maintaining or improving dosimetric quality[6]–[11].

Traditionally, treatment planning has been a manual, iterative process to be performed by human planners. In this process, planners repeatedly adjust treatment planning parameters (TPPs), such as objective weights, dose constraint values, etc., until a clinically acceptable treatment plan solution is found. The iterative planning process is not only time-consuming and labour intensive, but the resulting plan quality highly depends on planner skill and experience [12]–[19]. Automated methods, therefore, aim to reduce active planning time (with fully automated methods requiring no active planning).

In radiation therapy inverse planning, translating clinical goals into a multicriteria objective function for optimization is often very difficult [19]. In developing automated methods for treatment planning, we have three main considerations: (1) to

 Manuscript received xxxxxx; revised xxxxxx; accepted xxxxx. Date of publication xxxxxx; date of current version xxxxxx. This work was supported in part by the NIH (1R01 CA176553 and R01CA227713), National Institute of Biomedical Imaging and Bioengineering Training in Biomedical Imaging Instrumentation (NIBIB TBI2), and a Faculty Research Award from Google Inc.

C. Huang is with the Department of Bioengineering, Stanford University, Stanford, CA 94305.
S. Boyd is with the Department of Electrical Engineering, Stanford University, Stanford, CA 94305.
Y. Yang, N. Panjwani, and L. Xing are with the Department of Radiation Oncology, Stanford University, Stanford, CA 94305 (Ph: (650) 498-7896 E-mail: lei@stanford.edu).

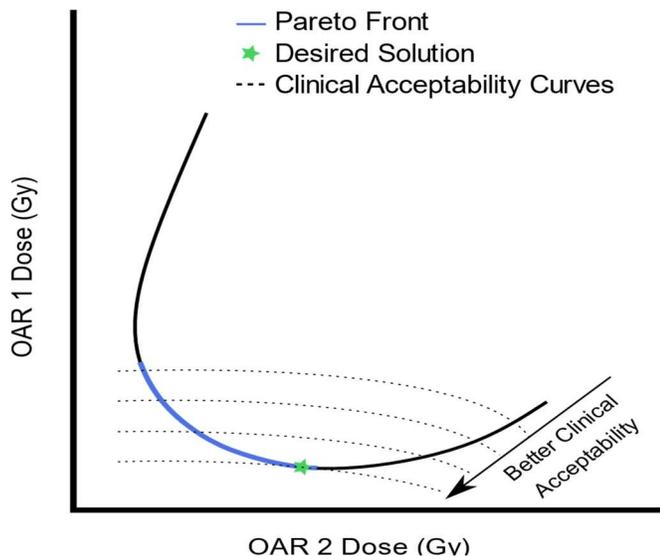

**Figure 1.** Visualization of the decision-making process for human planners in multicriteria optimization

reduce active planning time, (2) to produce acceptable plans that satisfy clinical criteria, and (3) to produce plans that are efficient.

From a design perspective, leaving performance on the table is highly undesirable. As treatment planning can involve many conflicting objectives (e.g. reduce mean OAR doses, minimize the deviation in PTV dose from the prescribed dose, etc.), no single plan can optimize performance on all objectives at once. The treatment planning problem can instead be approached using multicriteria optimization, with the goal of producing Pareto optimal, nondominated solutions. These Pareto optimal plans are efficient—that is we cannot improve one aspect (e.g., reduce the dose in one OAR) without compromising at least one other aspect (e.g., reduce the PTV dose) [20], [21]. Plans that are not Pareto optimal (i.e. dominated plans) are inefficient, and there exists more optimal plan(s) that, for instance, achieve better organ sparing for all OARs.

Pareto optimal planning enables planners to produce efficient treatment plans that can leverage a treatment planning system (TPS) configuration to the best of its abilities. Intuitively, producing efficient plans is a cornerstone of high-quality patient care, but not all Pareto optimal plans are acceptable clinically.

Figure 1 illustrates the decision-making process of human planners for selecting Pareto optimal plans based on improving clinical acceptability. Here, the blue curve represents the Pareto front, where not all plans are equally acceptable. The decision-making process of a planner might be conceptualized as selecting Pareto optimal plans that appear most appealing or clinically acceptable [21]. The goal of our work, therefore, is to automate this decision-making process and produce treatment plans that are both Pareto optimal and clinically acceptable.

*A. Related Works*

Here, we provide a summary of various automated treatment planning approaches and point to a more in-depth review[22], as well as the Supplemental Materials, for interested readers. In general, strategies of automating treatment planning can be categorized as knowledge-based planning (KBP) [23]–[26], protocol-based planning (PBP)[13], [14], [27]–[30], and multicriteria optimization (MCO) [12], [20], [31]–[35]. As each category of approaches has its own benefits and drawbacks, no clear consensus has been reached on an approach that can completely replace manual planning, and, in practice, a combination of various approaches may be used in a case-by-case fashion.

Our fully automated Pareto Optimal Projection Search (POPS) algorithm directly searches the Pareto front for clinically acceptable plans, doing so in a fully automatic fashion. Like previous MCO approaches, POPS produces Pareto optimal plans, ideally, without the limitations of *a priori* and *a posteriori* methods. Instead of a list of preferences based on general trends, POPS uses a detailed scoring function to evaluate plans on clinical acceptability. POPS additionally performs an automated search using such a scoring function, and, in theory, does not require any active planning from human planners as is necessary for *a posteriori* methods. Our results demonstrate that POPS indeed produces Pareto optimal treatment plans that perform at least as well as dosimetrist generated plans in terms of clinical acceptability.

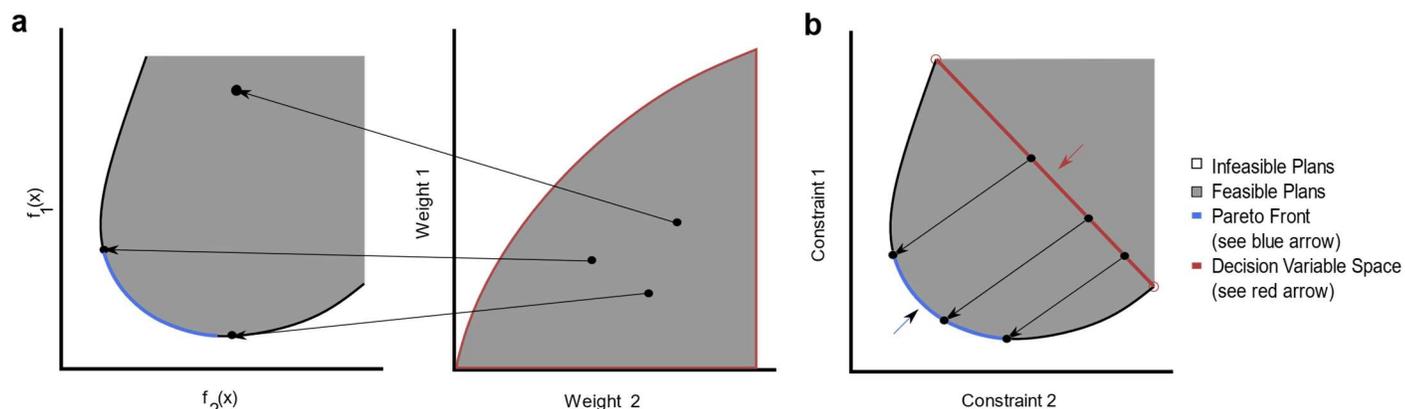

**Figure 2.** (a) Visualization of the separate decision variable and objective function spaces for weighted optimization problem setups. (b) Visualization for the direct projections between the decision variable space and the Pareto front for POPS

**Table 1.** List of clinical acceptability criteria used in scoring prostate patients, the priority of each OAR, and the alpha values for the piece-wise sigmoid function. Dose values are in units of Gy.

| | Clinical Acceptability Criteria | | | | | | Priority | $\alpha_1$ | $\alpha_2$ |
|---|---|---|---|---|---|---|---|---|---|
| Rectum | $D(80\%) \leq 30$ | $D(55\%) \leq 47$ | $D(40\%) \leq 65$ | $D(25\%) \leq 70$ | $D(10\%) \leq 75$ | $D(max) \leq 1.1D_p$ | 2 | 0.2 | 3 |
| Bladder | $D(80\%) \leq 30$ | $D(55\%) \leq 47$ | $D(30\%) \leq 70$ | | | $D(max) \leq 1.1D_p$ | 3 | 0.2 | 3 |
| Right Femoral Head | | | $D(max) \leq 50$ | | | | 4 | 0.2 | 3 |
| Left Femoral Head | | | $D(max) \leq 50$ | | | | 4 | 0.2 | 3 |
| Body | $D(20\%) \leq 8$ | $D(15\%) \leq 15$ | $D(10\%) \leq 25$ | | | $D(max) \leq 1.1D_p$ | 5 | 0.2 | 3 |

## II. METHODS

### A. Problem Formulation

The purpose of our fully automated POPS algorithm is two-fold: to produce treatment plans that are both Pareto optimal and optimal with respect to clinical acceptability. On one hand, alternative automated approaches like KBP and PBP do not directly address the issue of generating efficient plans. Alternative MCO approaches, on the other hand, discuss methods for generating the Pareto front while delegating the task of selecting the acceptable plan to a human planner. Our proposed approach hopes to further reduce active planning times for MCO methods by directly searching the Pareto front for clinically acceptable plans, producing plans that are both Pareto optimal and clinically acceptable.

As previously mentioned, Pareto optimal plans are non-dominated and further improvements to the plan in regards to one aspect are only made by trading-off in regards to other aspects. Intuitively, Pareto optimal treatment plans lie on the front between the region of feasible and infeasible plans. As described previously, one valid way of formulating the problem is to perform multicriteria optimization with multiple weighted objective functions. Such an approach, however, involves a separate decision space and objective function space, which contains the Pareto front (as depicted in Figure 2a). Because the decision variables (i.e. objective weights) in MCO do not typically have an intuitive mapping to the objectives, many MCO methods focus primarily on generating the Pareto front. Our proposed POPS algorithm instead formulates the iterative treatment planning problem as a feasibility search, which allows for a more direct relationship (projection) between the decision and constraint feasibility spaces (Figure 2b). We can then project points from the decision space to the constraint feasibility space and utilize any gradient-free searching algorithm to navigate to our desired treatment plan solution.

### B. Quantifying Clinical Acceptability

By convention, the evaluation of treatment plans has been a manual, iterative, and qualitative process. In performing their qualitative evaluation, physicians ideally draw on their repository of clinical experience to judge various components of the plan. Previous works that explore various DVH constraints have attempted to operationalize the considerations made in a physician's qualitative assessment [18], [19], [36], [37]. Similarly, evaluation of treatment plans through various metrics has been used previously in knowledge based planning [23]–[25], as well as more broadly for data analytics [18], [36], [37].

To summarize the general intuition that can be followed in quantifying clinical acceptability of treatment plans, the overall quality of a treatment plan can be determined by assessing its performance in regards to individual metrics and criteria. We can first define a list of criteria, termed clinical acceptability

$$SF = \begin{cases} 1, & \text{if solution is infeasible} \\ \dfrac{\sum_{s \in S} 2^{-Priority_s+1} \cdot \dfrac{\sum_{i \in s} \sigma_i^{\pm}(PV_i - CV_i, \alpha_1, \alpha_2)}{N_s}}{\sum_{s \in S} 2^{-Priority_s+1}}, & \text{otherwise} \end{cases} \quad (1)$$

$$\sigma_i^{\pm}(z_i, \alpha_1, \alpha_2) = \begin{cases} \dfrac{1}{1+e^{-\alpha_1 z}}, & z \leq 0 \\ \dfrac{1}{1+e^{-\alpha_2 z}}, & z > 0 \end{cases} \quad (2)$$

criteria, that a physician might incorporate into their evaluation. As a foundation for our proposed list, we incorporate criteria from the list of DVH constraints discussed in Chen et al.[38]. While this particular list of criteria may not be agreeable to every physician, the entries can be readily changed to suit individual preferences or to follow various institutional protocol. Essentially each criterion represents a chosen control point on the DVH, and while we use the control points identified in Chen et al., other control points may certainly be used to similar effect. The main heuristic we follow in selecting these criteria is to provide relatively uniform sampling to the DVH for each organ. Based on advice from three consulted physicians and our uniform sampling heuristic, we then incorporate additional criteria to judge sparing of the body, separate from sparing of surrounding OARs. The full list of criteria is included in Table 1. Here, $D(\cdot \%)$ refers to dose at a given percent volume, $D(max)$ refers to the maximum dose, and $D_p$ refers to the prescription dose. Each criterion can be conceptualized as a bound on the dose received by a certain OAR volume. The first rectum criteria ($D(80\%) \leq 30$) could be understood as "the dose received by 80% of the volume should be at most 30 Gy."

From a list of criteria, we can then design individual scoring functions ideally tailored to the criteria and structure being judged. Previous works have proposed a variety of scoring functions [18], [36], [37], [39]. Some of these scoring functions (e.g. step function, regret, etc.) are too broad and lack organ-specific considerations [18], [37]. Others, like the generalized evaluation metric (GEM) [36], attempt to incorporate organ-specific scoring at the cost of maintaining a lengthy history of treatment plan solutions. As defined in the GEM score, organ-specific scoring is represented by a gamma distribution probability term. To maintain an accurate gamma distribution, the GEM score requires a lengthy history of treatment plans. However, it is preferable to maintain organ-specific scoring without requiring a lengthy history, so we modify the previously proposed GEM score by replacing its gamma distribution probability term with piece-wise sigmoid functions. Like the original GEM score, our proposed scoring function (SF) outputs a value between 0 and 1 (lower is better) where a score of 0.5 represents a plan that satisfies all clinical acceptability criteria, on average.

As shown in Equation 1 and 2, the intuition of organ-specific scoring is encapsulated by tailored piece-wise sigmoid functions (where $\alpha_1$ and $\alpha_2$ describe the steepness of the sigmoid functions and are chosen empirically) and a priority-based weighting (where the weighting scheme is adapted from Mayo et al. [36]). To select the steepness of the sigmoid functions, we adhere to the following heuristics:

1. Decreasing DVH values for each organ imply better sparing to that organ
2. Organ doses approaching 0 have diminishing returns, in regards to score
3. Not satisfying the listed clinical acceptability criteria is highly undesirable and intuitively results in a bad score

Following these heuristics, $\alpha_1$ is selected to be small in order to abide by heuristic 2 and $\alpha_2$ is selected to be large relative to $\alpha_1$ in order to abide by heuristic 3. $\alpha_1, \alpha_2 \in (0, \infty)$, and α values on the lower and upper ends of that interval can be interpreted as the sigmoid function being a flat line and step function, respectively. Priority values for each OAR were selected based on advice from three consulted physicians.

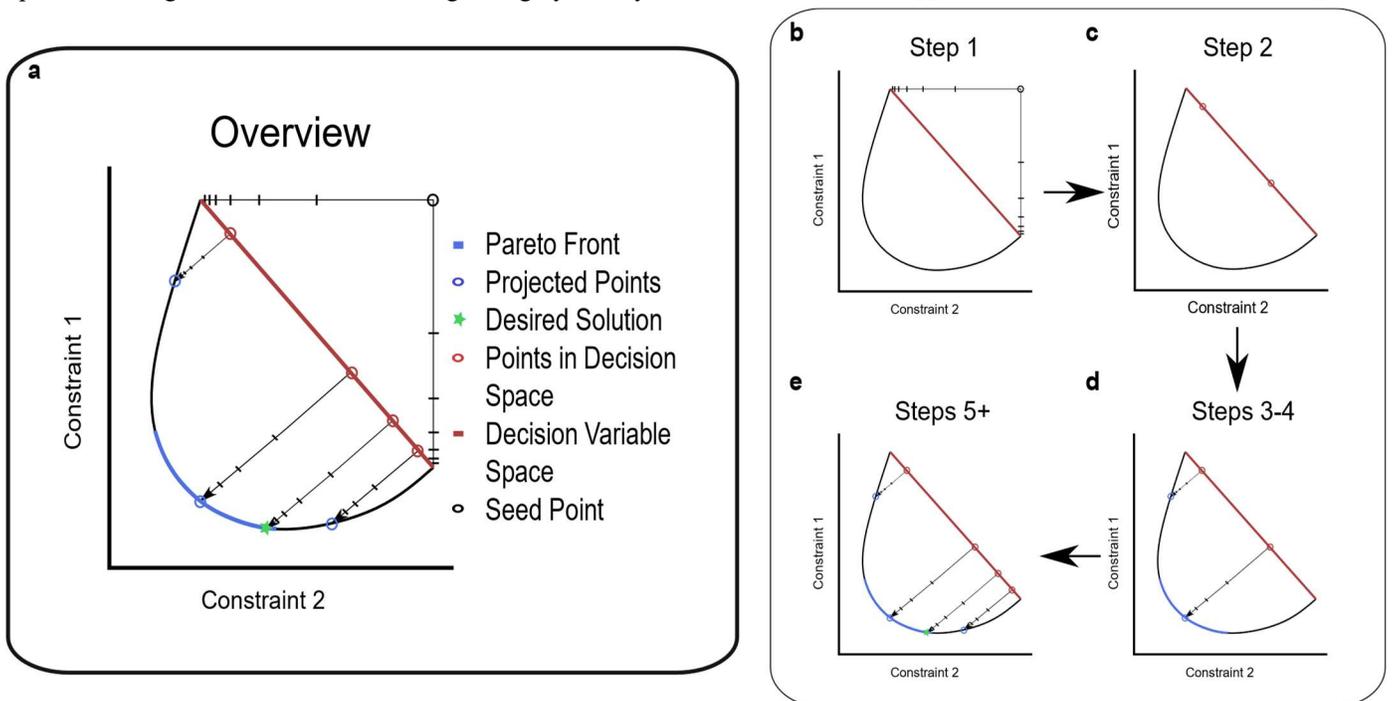

**Figure 3.** (a) Visualization of the POPS algorithm for a hypothetical 2D problem. (b) Starting from a seed point, we define the bounds for the decision variable space by projecting the seed point using the bisection method. (c) We then define the initial simplex and (d) project the initial simplex to the Pareto front, while computing the SF score for each projected point. (e) We can then use a gradient-free search (i.e. simplex search) to search the Pareto front for a clinically acceptable treatment plan.

Following the precedent of Mayo et al.[36], our clinical acceptability criteria do not include the target volume. For prostate IMRT, where the target volume is relatively large and the dose distribution to the target is relatively homogeneous (at least for the dose constraints used here), inclusion of target volume clinical acceptability criteria may be unnecessary. For other regions of the body or in particularly complex treatment planning cases, target volume criteria can certainly be incorporated without affecting POPS performance.

In Equation 2, $s$ refers to a structure (OAR) in the structure set $S$. The priority of a specific structure, $Priority_s$, adjusts its exponential weighting. $i$ refers to a clinical acceptability criteria (as listed in Table 1), $N_s$ refers to the total number of clinical acceptability criteria for a specific structure, and $\sigma^{\pm}$ refers to the structure score computed using a piecewise sigmoid function of the difference between the plan value $PV_i$ and the criteria value $CV_i$.

As an example, to compute the structure score for the second bladder criteria ($D(55\%) \leq 47$), we first compute the difference $z_i = PV_i - CV_i$. Here, $CV_i = 47\ Gy$ and $PV_i$ is found as the corresponding dose on the bladder DVH for 55% volume. In this toy example, we use a plan value $PV_i = 20\ Gy$, $\alpha_1 = 0.2$, and $\alpha_2 = 3$. We would then compute the structure score $\sigma_i^{\pm}(z_i, \alpha_1, \alpha_2) = 0.0045$. Assuming the solution is feasible, the scoring function then performs a weighted-average using OAR priorities. Our results use the priorities as listed in Table 1, which were determined based on advice from three consulted physicians, but they may also be changed to better suit individual planner preferences.

Intuitively, our proposed SF scores treatment plans on a scale between 0 and 1, where lower scores are better. Plans that attain scores of 0.5 satisfy the listed clinical acceptability criteria, on average. Plans that attain scores smaller than 0.5 achieve better OAR sparing, on average, than a plan that just satisfies the listed clinical acceptability criteria.

*C. POPS Algorithm*

The general methodology of our proposed POPS algorithm combines two search methodologies: 1) a projection from the decision variable space to the Pareto front using the bisection method [40] and 2) a gradient-free search (i.e. simplex search) of the decision variable space—and the corresponding points on the Pareto front—for a clinically acceptable treatment plan. Our implementation utilizes Nelder-Mead simplex search [41] due to its simplicity, but our methods can incorporate any gradient-free optimization approach. To allow for better reproducibility of our results, we implement our approach using the open-source MatRad software package [4].

*1) Seed Point Selection*

To start, we first select a seed point, which will later be used to bound the decision variable search space. Seed point selection varies with the treatment region (e.g. prostate, head and neck, lung, etc.), as each region has differing OARs each with differing dose constraints.

For our experiments on prostate cases, we chose to set the seed point coordinate to the maximum allowed EUD value of each OAR. Setting the seed point too small can cause the algorithm to miss portions of the Pareto front, as they will be outside the bounds of the decision variable search space. Conversely, setting the seed point arbitrarily large would lead to longer computation times, as the bisection method will need to be run over more iterations to project points to the Pareto front.

*2) Projection Method*

In conventional MCO approaches with weighted optimization, the decision variable search space (i.e. objective weights) and Pareto front can be present in separate spaces (see Figure 2). Whereas in the POPS feasibility search problem, we define the decision variable search space and the Pareto front to be in the same coordinate space (see Figure 2). In doing so, we can also define a mapping between each point in the decision variable search space and a corresponding point on the feasibility boundary by conducting a feasibility search using the bisection method. For conciseness, we can refer to this feasibility search as a projection of points in the decision variable search space to the feasibility boundary.

In our implementation, we perform projections by conducting a one-dimensional search using the bisection method [40] along a vector direction (see the segmented vectors in Figure 3). In this one-dimensional search using the bisection method, we begin with a starting point and search along a vector direction for the feasibility boundary. At each iteration of this feasibility search, we compute the midpoint of the current interval (at the beginning this interval is defined by the start and end points) and evaluate whether the midpoint is a feasible plan. If the bisection method has not converged, we then select a new interval—either as the start point and midpoint or the midpoint and end point—and continue the feasibility search.

We define the coordinates of a point $p$ in the decision variable space as $(c_1, c_2, \dots, c_n)$ (i.e. red circles in Figure 3), the

$$
\begin{aligned}
\min_x \quad & \frac{1}{N_{ptv}} \sum_{j \in ptv} (d_i - D_p)^2 \\
s.t. \quad & x \geq 0 \\
& \vec{d} = D\vec{x} \\
& c_{EUD,rectum} \leq c_1 \\
& c_{EUD,bladder} \leq c_2 \\
& c_{EUD,FH\ R} \leq c_3 \\
& c_{EUD,FH\ L} \leq c_4 \\
& c_{EUD,body} \leq c_5 \\
& D_{ptv}(95\%) \geq D_p \\
& D_{ptv}(min) = 0.97 D_p \\
& D_{ptv}(max) = 1.1 D_p
\end{aligned}
\qquad (3)
$$

coordinates of a projected point $p'$ on the Pareto front as $(c'_1, c'_2, ..., c'_n)$ (i.e. blue circles in Figure 3), and the SF score of the projected point as $f(p')$. We can then formulate the iterative treatment planning problem as a feasibility search using Equation 3. We note that POPS performs equally well if other constraints are chosen besides equivalent uniform dose (EUD) (i.e. mean, max, and DVH constraints). In this feasibility search, we minimize the placeholder objective of the squared difference in dose between PTV voxels and the prescription dose $D_p$. The values $c_1, ..., c_5$ can be conceptualized as bounds on the EUD constraints for various OARs.

For clarity, we provide visualizations and descriptions for a hypothetical 2D case below (see Figures 2 and 3 for visualizations of the Pareto front in 2D). However, our POPS algorithm generalizes to n-dimensional cases, and results are provided in later sections for 5D prostate IMRT cases. The algorithm implementation details and pseudocode are provided in the Supplemental Materials.

## III. RESULTS

### A. Experimental Setup and Evaluation

To determine the proficiency of our automated POPS algorithm, we compare POPS generated IMRT treatment plans to gold-standard VMAT treatment plans created as part of routine clinical workflow. While these human-created treatment plans are not necessarily Pareto optimal, they provide a benchmarking baseline in terms of clinical acceptability. Currently, treatment plan evaluation does not have a widely adopted evaluation metric for comparing treatment plans in regards to clinical acceptability. Instead, it is common to perform quantitative analysis of a plan's statistics.

To perform treatment plan evaluation, it is important to evaluate plans based on dose conformity, dose homogeneity, and OAR sparing [30], [42]–[48]. We adopt the following definition of the conformity index (CI) [48] and homogeneity index (HI) [47] to assess dose conformity and homogeneity, respectively:

$$CI = \frac{\left(TV_{95D_p}\right)^2}{TV \times V_{95D_p}} \quad (4)$$

$$HI = \frac{D_5 - D_{95}}{D_p} \quad (7)$$

In order to assess OAR sparing, we compute the DVH values at five uniformly sampled control points (D(20%), D(40%), D(60%), D(80%), D(98%)). Lower dosage values at each of these control points implies better OAR sparing. We additionally conduct a Wilcoxon signed-rank test comparing plan statistics.

The dataset used in our experiment consists of 21 prostate cases acquired from the clinic following Stanford IRB protocol #41335. Scanner details, acquisition dates, and treating physician varied across cases. Gold-standard VMAT treatment plans used two full coplanar arcs and were created by a human planner using a treatment planning system from Varian Medical Systems. Various OARs (including the rectum, bladder, left/right femoral heads, and body) were contoured, along with the PTV. In order to mitigate the effect of beam angle selection on plan quality and help control for dosimetric differences between IMRT and VMAT, we use a plan setting of 9 equally spaced photon beams (from 0° to 360° in increments of 40°). All POPS plans used a prescription dose of 74 Gy delivered over 40 fractions, a bixel size of 5 $mm$, and a dose voxel size of $3 \times 3 \times 3\ mm^3$. Our POPS algorithm additionally implements the open-source MatRad software package to perform inverse planning [4] which uses a pencil-beam dose calculation algorithm and IPOPT[49].

Note that prior works have shown that prostate VMAT plans tend to have better dosimetric quality than IMRT plans [9]–

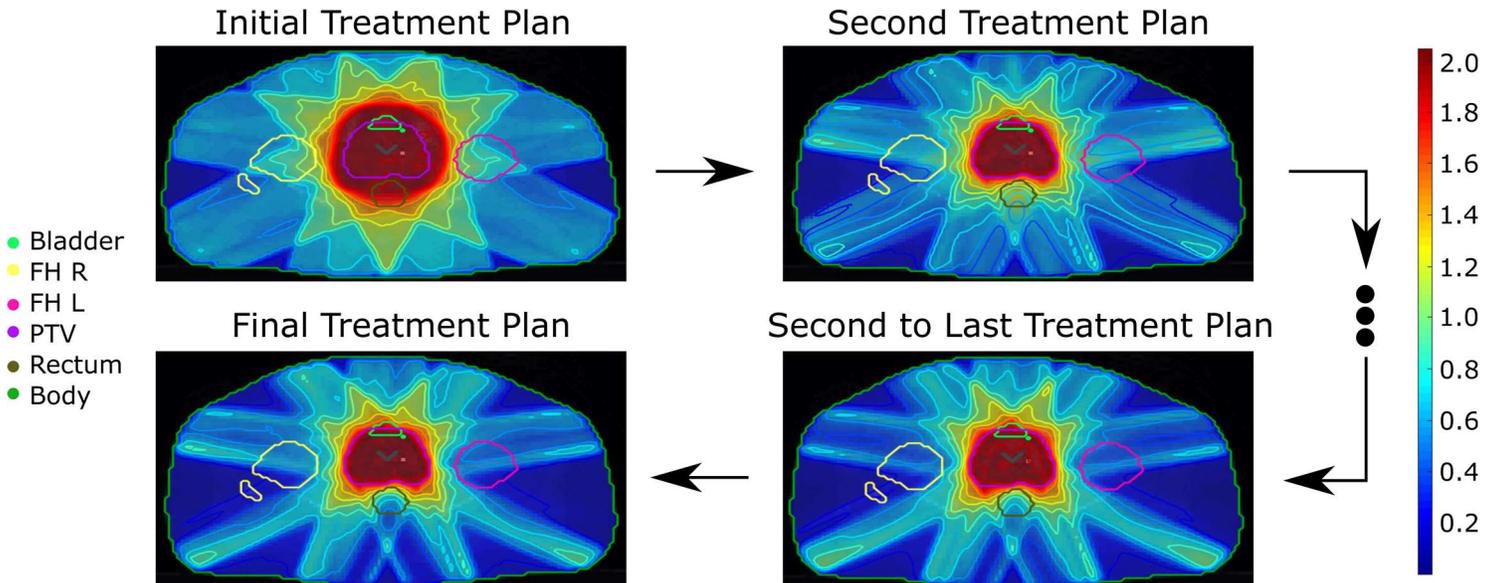

**Figure 4**. Visual comparison of initial, intermediate, and final treatment plans generated for an example patient using POPS. Dose conformity and organ sparing significantly improve from the initial plan after running POPS.

[11]. However, we compare POPS results for IMRT to results for VMAT planning as we do not have access to a more relevant clinical IMRT dataset. VMAT planning, moreover, represents a current gold-standard in the clinic, and we plan to extend POPS to VMAT planning in future work.

### B. Qualitative Comparison

We first conduct a qualitative comparison between plans produced by POPS and a human planner. A visualization of the differences in dose distribution between an example of a representative initial plan and the corresponding POPS generated plan is provided in Figure 4. We can visually appreciate the improvement to plan quality when comparing the final and initial plans. Moreover, we can visualize incremental improvements following iterations of the POPS algorithm. Upon arrival at the final plan, we observe substantial improvements to OAR sparing for all OARs as compared to the initial plan, as well as incremental improvement to sparing of the rectum when comparing the final plan to intermediate plans. Dose delivered to the PTV remains relatively homogenous throughout all iterations with dose conformity being noticeably improved in the final plan when comparing to the initial plan.

Figure 5 provides visualizations of three example cases sorted by SF score: 1) where the POPS provided a large improvement over the dosimetrist generated plan, 2) where POPS provided a moderate improvement, and 3) a case where the POPS and dosimetrist generated plans scored similarly. It is clear from all three cases that POPS generated plans perform better in sparing the rectum, which is assigned the highest priority in our SF score following the advice of three consulted physicians.

In the DVH for the large improvement case (row 2 column 1 of Figure 5), we can appreciate substantial improvement to the DVH curves for the rectum (purple line) and femoral heads (green lines), along with similar performing DVH curves for the bladder and body. Here, the dosimetrist generated plan may not be Pareto optimal. In these scenarios, POPS produces more efficient plans than the dosimetrist, highlighting the importance of Pareto optimality in treatment plan quality. The dose distribution (row 1 column 1 of Figure 5) for this large improvement case additionally demonstrates that PTV dose is highly conformal and the algorithm performs well in OAR sparing. In the DVH for the moderate improvement case (row 2 column 2 of Figure 5), we observe significant improvements over the dosimetrist plan in rectum sparing (purple line) at the expense of sparing to bladder (blue line). We suspect this trade-off occurs because the dosimetrist plan is close to Pareto optimal, and POPS—with our current scoring function—assigns greater priority to the rectum than the bladder. Finally, in cases where POPS generated plans and dosimetrist generated plans score similarly, it is likely that the dosimetrist generated plans are close to Pareto optimal, and the dosimetrist's assignment of OAR priority agrees with the OAR priorities listed in Table 1.

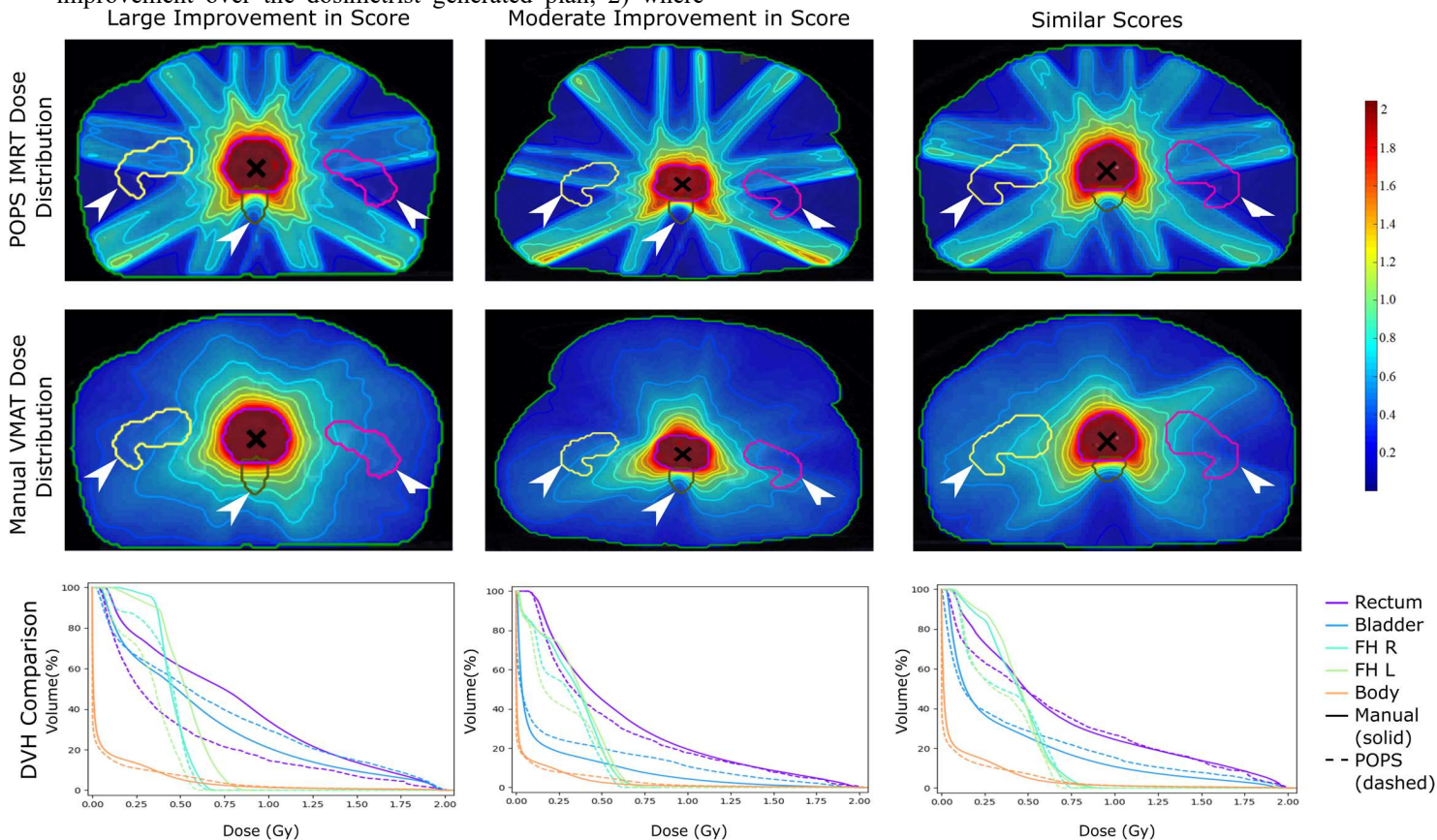

**Figure 5.** Visualization for three representative patients where POPS provided (1) large improvement, (2) moderate improvement, and (3) a similar SF score to physician generated plans. A visualization of the POPS generated dose distribution, manual VMAT dose distribution, and DVH comparisons are provided for each case. Arrows indicate regions where the POPS dose distribution has better OAR sparing. See the end of Section IIIA for an explanation on the visual differences in dose distributions.

*C. Quantitative Comparison*

Tables 2 and 3 provide further quantitative comparison between dosimetrist generated VMAT plans and POPS generated IMRT plans. To evaluate OAR sparing, five DVH control points (D(20%), D(40%), D(60%), D(80%), D(98%)) were selected to provide an approximation of the DVH curve for each OAR. Similarly, we evaluate dose conformity and homogeneity using the CI and HI, respectively. All differences were quantified using the Wilcoxon signed-rank test ($p < 0.05$), and p-values were corrected using the Holm-Bonferroni method to control family-wise error rate [50].

Dose conformity values are summarized in Table 2, where CI approaches 1 for an ideal case. Plans produced using POPS and those produced by a human planner had comparable dose conformity. CI values for both POPS plans and gold-standard plans are within an acceptable range and comparable to values reported in previous work for prostate VMAT cases [44].

Dose homogeneity values are also provided in Table 2, where smaller HI values are better. Here, gold-standard VMAT plans provided more homogeneous dose distributions than those for POPS IMRT plans. Both POPS and the gold-standard plans had better HI values than reported in previous work [44], and both were within a clinically acceptable range.

Similarly, dosage values are comparable between the two methods, demonstrating comparable OAR sparing. While POPS generally performs better for the rectum, femoral heads, and body, it performs slightly worse for the bladder, which may be due to a different emphasis on the bladder in manual plans.

**Table 2.** Comparison of CI and HI for POPS IMRT plans and gold-standard VMAT plans created manually. CI values approaching 1 are better, and smaller HI values are better.

|  | Conformity Index (CI) | Homogeneity Index (HI) |
|---|---|---|
| Manual VMAT | 0.862 (0.026) | **0.046 (0.010)** |
| POPS IMRT | 0.831 (0.051) | 0.104 (0.001) |
| Wilcoxon signed-rank test p-value | 0.08503 | **0.00214** |

Modifying the emphasis on the bladder by changing its weighting in the scoring function could help to reduce this discrepancy and would be a possible avenue for future work. Of the 25 DVH control points evaluated here, POPS plans had significantly better OAR dosages for 13 control points, worse dosages for 2 control points, and similar dosages for 8 control points. Overall, POPS produces Pareto optimal plans that are highly comparable to dosimetrist generated plans with regards to clinical acceptability.

## IV. DISCUSSION

This study introduced a fully automated treatment planning algorithm, POPS. POPS combines projections using the bisection method and a gradient-free search to produce treatment plans that are both Pareto optimal and clinically acceptable. To evaluate clinical acceptability, we compare performance for POPS IMRT planning to performance for VMAT planning by dosimetrists, the current gold-standard.

*A. POPS vs Other Approaches*

Automated treatment planning approaches have steadily grown in popularity due to their potential to drastically reduce

**Table 3.** Comparison of DVH values for five uniformly selected control points (D(20%), D(40%), D(60%), D(80%), D(98%)) between dosimetrist generated and POPS generated plans. Lower dosage values are better as they imply better OAR sparing. The significantly lower dosage values and significant p-values are bolded (significance when $p < 0.05$).

|  | OAR | D(20%) (Gy) | D(40%) (Gy) | D(60%) (Gy) | D(80%) (Gy) | D(98%) (Gy) |
|---|---|---|---|---|---|---|
| Manual VMAT | Rectum | 52.5 (10.9) | 32.0 (7.9) | 20.6 (7.1) | 9.7 (5.1) | 3.0 (1.1) |
|  | Bladder | **35.7 (21.2)** | **14.8 (11.7)** | 6.8 (7.0) | 3.8 (5.0) | 2.0 (2.6) |
|  | FH R | 24.2 (4.8) | 19.8 (4.3) | 14.8 (4.0) | 7.3 (4.3) | 2.2 (3.0) |
|  | FH L | 22.4 (4.6) | 17.4 (3.3) | 13.4 (3.2) | 6.9 (4.3) | 2.1 (2.9) |
|  | Body | 3.4 (1.7) | 0.6 (0.2) | 0.2 (0.1) | 0.1 (0.0) | 0* |
| POPS IMRT | Rectum | 48.7 (10.7) | 27.0 (8.7) | **13.7 (4.8)** | **6.0 (2.5)** | **2.3 (1.3)** |
|  | Bladder | 46.5 (18.6) | 24.3 (15.9) | 10.9 (8.5) | 4.9 (4.9) | 2.1 (3.0) |
|  | FH R | 20.8 (4.3) | **15.0 (5.3)** | **7.9 (4.3)** | **4.3 (2.1)** | **1.2 (1.0)** |
|  | FH L | 20.3 (4.1) | **13.5 (4.6)** | **7.2 (3.8)** | **3.6 (1.3)** | **1.1 (1.0)** |
|  | Body | 3.2 (1.9) | **0.2 (0.5)** | **0.1 (0.0)** | 0* | 0* |
| Wilcoxon signed-rank test p-value | Rectum | 0.13653 | 0.09424 | **0.01174** | **0.00353** | **0.01274** |
|  | Bladder | **0.00214** | 0.00214 | 0.09064 | 0.87571 | 0.39163 |
|  | FH R | 0.06271 | **0.01823** | **0.00448** | **0.00618** | **0.02378** |
|  | FH L | 0.18262 | **0.01274** | **0.00214** | **0.00214** | **0.01866** |
|  | Body | 0.39163 | **0.01328** | **0.01328** |  |  |

*Values were vanishingly small

active planning time [13], [14], [23], [27], [28], [30], [32], [51], [52]. Following the iterative approach to treatment planning, human planners repeatedly adjust treatment planning parameters and perform inverse planning until a clinically acceptable solution is found. However, an acceptable treatment plan that just satisfies clinical dosimetric criteria is not necessarily the most efficient, and perhaps plan parameters can be tuned further until a planner finds the best achievable plan. The decision-making process of treatment planners could be interpreted as optimizing treatment plans for both acceptability and efficiency. This process is often both time-consuming and labour intensive for planners. For IMRT cases, active planning time—*planning time that directly utilizes a human planner's decisions or actions*—has been reported to be on the order of 135 minutes, with active planning time potentially increasing for more complex cases [32].

Endeavouring to reduce active planning time, a variety of automated approaches have been proposed in literature, which we introduced in earlier sections. Of these various approaches, MCO methods attempt to produce Pareto optimal solutions. Where *a posteriori* MCO (database MCO) approaches attempt to generate the entire Pareto front (or at least approximate the Pareto front) [20], [31], [32], *a priori* MCO approaches select a Pareto optimal plan based on physician-defined preferences for objective weights and dose constraints[12], [33], [52], [53]. Both types of MCO approaches produce Pareto optimal plans but delegate the task of selecting clinically acceptable plans to the physician, either explicitly (i.e. in the case of *a posteriori* MCO) or implicitly through preferences (i.e. in the case of *a priori* MCO).

While we do not have access to the implementations of other MCO algorithms in the MatRad framework, we report their active planning times for rough comparison in Table 4. Our initial implementation of POPS requires no active planning from human planners. To the best of our knowledge, POPS is currently the only approach capable of producing Pareto optimal and clinically acceptable plans with no active planning required. Alternative *a posteriori* MCO approaches, for reference, require a human planner to select clinically acceptable plans from a database of Pareto optimal solutions, resulting in longer active planning times compared those for POPS.

One major benefit of having no active planning time is that many patients can be run in parallel, reducing the average time overall per patient and allowing for great scalability to servers or clusters. For our 5D prostate IMRT implementation, POPS utilizes 5 CPU threads. On our consumer-level desktop Ryzen 2700x CPU (16 threads), up to 3 patients may be run simultaneously, which scales to potentially 25 patients on a Ryzen 3990x workstation and even more when factoring in servers or a gpu-based implementation. Similarly, having no active planning requirement allows physicians and dosimetrists to better allocate their time in the clinic, potentially spending it on more demanding and relevant tasks.

Overall, our proposed POPS algorithm directly searches the Pareto front for clinically acceptable treatment plans, as defined by a scoring function. We note that the particular scoring function can be interchanged, as desired, without affecting the functionality of the POPS algorithm. In contrast to previous categories of MCO approaches, POPS automates the selection process of a clinically acceptable plan. In practice, POPS provides a general framework that allows for the direct search of the Pareto front. Based on results for 21 prostate cases, POPS IMRT plans have comparable dose conformity and OAR sparing with slightly worse dose homogeneity, as compared to VMAT plans produced in clinical workflow.

The proposed scoring function used by POPS incorporates clinical acceptability criteria (Table 1) proposed in previous work [38]. As treatment plan evaluation can be subjective, we certainly acknowledge that physicians can choose criteria that may be somewhat different from what we adopted in the present study. To that end, Table 1 can be modified as desired to better suit individual preferences without affecting the performance of the POPS algorithm. Similarly, we found our heuristics to work well for prostate IMRT planning, but they may also be modified, resulting in different hyperparameter values, to suit different anatomies or physician preferences. The POPS framework allows for the implementation of alternative scoring functions or clinical acceptability criteria. For a given scoring function and set of criteria, POPS produces treatment plans that satisfy the overall objective of this study: to produce treatment

**Table 4.** Comparison of average computation times and active planning times required for each patient. As compared to traditional MCO approaches, POPS trades active planning time for background computation time, allowing dosimetrists and physicians to better allocate their time in the clinic. For this study, we define active planning time as planning time that directly utilizes a human planner's decisions or actions.

| Method | Active Planning Time/Patient | Computation Time/Patient | Total Time/Patient |
| --- | --- | --- | --- |
| Human Planner[*] | 135 min | N/A | 135 min |
| Approximated MCO (with RayStation implementation)[*] | 12 min | 7 min | 19 min |
| POPS (with MatRad implementation) | 0 min | 78 min | 78 min |

[*] Based on reported average time from Craft et al. [32]

plans that are both Pareto optimal and clinically acceptable in a fully automated fashion.

*B. Limitations and Potential Improvements to Speed*

Based on the reported computation times from Craft et al., MCO approaches that utilize approximations of the Pareto front typically have computation times less than 10 minutes. Our current implementation, which uses the MatRad software package to perform inverse planning and does not yet utilize Pareto front approximations, has computation times around an hour. We would like to clarify that the computation time differences are not the result of POPS. Rather, they can be attributed to two main factors: inverse planning software implementation differences and the PGEN approximation method (or equivalent Pareto front approximation method). Each time POPS makes an adjustment to the search variables, the vast majority of time is spent in computing function evaluations (i.e. performing inverse planning), so the bottleneck to sequential throughput is inverse planning speed. We have implemented our POPS algorithm using the open source MatRad package so that our findings can be more easily verified by other studies, but implementation of POPS in alternative treatment planning software packages like Eclipse or RayStation or ConRad [3] will speed up computation accordingly.

For our implementation, we make no additional assumptions or approximations in regards to dose constraints or objectives (beyond those made by the matRad software package), but recent works such as ConRad demonstrate dramatic improvements to inverse planning throughput if certain approximations are made [3]. Similarly, as reported by Craft et al. [31], [32], their implementation uses the PGEN approximation of the Pareto front, requiring significantly fewer function evaluations to approximate the Pareto front. We hope to apply both improvements to our POPS algorithm in the future and anticipate similarly large increases in speed as were observed in their works.

V. CONCLUSION

External beam radiation therapy is used for treatment of a significant number of cancer patients[54]. Within the radiation therapy workflow, the treatment planning process represents a bottleneck to high quality patient care, due to the time-consuming nature of the iterative planning process and inter-planner variability. Previous works in automated planning can be generally categorized as KBP, PBP, and MCO approaches, each having key limitations. Our proposed POPS algorithm attempts to address these limitations and provides a fully automated framework for performing Pareto optimal treatment planning.

Evaluation of treatment plan quality is an especially subjective and nuanced process. While we acknowledge that particular scoring of treatment plans may differ according to planner expertise and preferences, we adopt a scoring function (SF) for the purpose of optimizing prostate treatment plan quality in regards to clinical acceptability criteria previously proposed in literature [38]. Through qualitative and quantitative analyses comparing IMRT plans produced by POPS with gold-standard VMAT plans produced as part of clinical workflow, we demonstrate that our proposed POPS algorithm produces plans that are both Pareto optimal and clinically acceptable. While our initial implementation of POPS has longer computation times than alternative approaches, POPS is currently the only MCO-like approach that has no active planning time requirement. Overall, our results indicate that POPS can substantially improve treatment planning workflow.

VI. ACKNOWLEDGEMENT

The authors would like to thank Varun Vasudevan for his help in pre-processing and curating treatment plan data collected as part of clinical workflow. The authors would also like to thank and acknowledge the MatRad development team for their help and advice.

# Supplemental Materials

## I. INTRODUCTION

### A. Related Works (continued)

KBP methods are a category of methods based on the premise that treatment planning results can be predicted from the geometric and anatomical information of a patient [23]. KBP typically follows the paradigm of training a supervised machine learning model on a dataset of previously generated treatment plans. Given input information in the form of a patient's CT and structure segmentations, a knowledge-based model attempts to predict voxel-wise dose or the dose-volume histogram (DVH) [23]–[26]. Previous works report that knowledge-based models can accurately predict treatment planning results [23]–[25], but these results are often limited to a subset of the information in treatment plans (i.e. dose distribution, DVH, etc.) and are not equivalent to performing the treatment planning process. In cases where KBP performs poorly, lack of model interpretability limits the usability of KBP approaches. Moreover, because KBP is typically formulated as a supervised learning problem, it makes no guarantees about the efficiency of produced plans and, instead, only attempts to produce plans that are similar to the ground-truth data used in training, which may be a limiting factor on algorithm performance. Combined, these drawbacks may limit the practical usability of KBP methods in the clinic, at least in their present forms [55].

PBP methods are a category of methods that employ rules or heuristics to mimic an experienced treatment planner performing iterative planning. Previous works have proposed various rules for adjusting treatment planning parameters (i.e. objective weights, constraint values, etc.) based on advice from consulted physicians [14], [27]–[30]. Others have followed a model-free approach, opting to use reinforcement learning to mimic human planners [13].

While protocol-based methods may offer a practical approach to automation, they do not directly address the issue of inter-planner variability. In particular, the design of rules or heuristics for protocol-based planning highly depends on the perspective of the planner, leading to variable plan quality between institutions or even individual planners[14], [15]. Automated approaches, ideally, should remove inter-planner variability and provide a baseline method for producing high-quality treatment plans, but the issue of inter-planner variability remains very present in protocol-based approaches.

Unlike KBP and PBP approaches, MCO approaches attempt to generate Pareto optimal treatment plans. MCO approaches can be further categorized into *a posteriori* MCO [20], [31], [32] and *a priori* MCO [12], [33], [34].

In *a priori* MCO, a Pareto optimal solution is found according to provided preferences [19], [51]. Such approaches require sufficient preference information to be provided by the planner, and some notable examples include scalarization (e.g. $\epsilon$ - constraint method, achievement scalarization, etc.) [52], prioritized optimization [12], [33], [34], and the lexicographic method [53]. Translating physicians' intuitions regarding ideal planning to a list of preferences, however, is no trivial task, and sometimes treatment plan solutions found from preference information may not align with a planner's expectations. Where preference information may only capture general trends observed by a physician, more nuanced metrics that utilize dose-volume histogram information is likely necessary.

*A posteriori* MCO, by contrast, explicitly place the decision of selecting treatment plans in the hands of the planner. These methods attempt to generate or approximate the Pareto surface, creating a database of Pareto optimal treatment plans that a physician might choose from. Previous works have proposed alternative ways to generate the Pareto surface (e.g. Pareto surface generation for convex multi-objective instances or PGEN, simulated annealing, evolutionary algorithms, etc.) or at least approximate its shape assuming a convex formulation to the problem [20], [31], [32]. *A posteriori* methods, in general, produce Pareto optimal and clinically acceptable plans, though at high computational cost. Further, these methods still require physicians to select clinically acceptable plans from the generated database and, therefore, are not fully automated methods.

## II. METHODS

### C. POPS Algorithm

#### 3) POPS Algorithm Implementation

*Step 1. Define the bounds $p_{b,1}, p_{b,2}, \ldots, p_{b,n}$ for the decision variable search space by projecting a seed point $p_s$ onto the Pareto front.*

When projecting the seed point to define the decision space bounds, we perform the one-dimensional search by tightening the EUD constraint in one OAR (while fixing all other constraints) and repeat this projection for each OAR (five times total for a 5D prostate case). The projected points will form the bounds for the decision space and can be visualized as a simplex (see Figure 3b).

As visualizations for n-D are too difficult, we instead provide visualizations in 2D where the feasibility search space is in $R^2$ while the decision space is the line bounded by the initial projected points. For a n-dimension treatment planning problem, the constraint feasibility search space is in $R^n$ while the decision space is a simplex bounded by $n$ points that are found by projecting the seed point along orthogonal directions.

*Step 2. Within the bounded decision space, define the initial simplex vertices $p_1, p_2, \ldots, p_n$.*

For simplicity, we define our initial vertices by taking the weighted average of the boundary points.

$$\tau_{i,j} = \begin{cases} 0.3, & \text{if } i = j \\ 1, & \text{otherwise} \end{cases} \quad (5)$$

$$p_i = \frac{\sum_{j=1}^{n} p_{b,j} \tau_{i,j}}{\sum_{j=1}^{n} \tau_{i,j}} \quad (6)$$

Here, $\tau_{i,j}$ refers to the weights assigned to each boundary point $p_{b,j}$. Prior knowledge can also be incorporated in the form of better starting simplex vertices, but our implementation uses a more straightforward initialization for reproducibility.

*Step 3. Project each of the simplex vertices $p_1, p_2, \ldots, p_n$ from the decision space to the Pareto front to obtain the projected points $p'_1, p'_2, \ldots, p'_n$, again using the bisection method.*

Projection of the simplex vertices to the Pareto front is visualized for the 2D case in Figure 3d. Similar to the projection of the seed point, all subsequent projections use the bisection method [40] to perform a one-dimensional search along a fixed vector direction. To determine this vector direction, we empirically chose to calculate the vector from the centroid of the initial simplex to the origin. Alternative choices to calculate the vector direction, such as choosing the normal vector to the decision variable plane, can work just as well. Example visualizations for one-dimensional search along this vector direction are shown as solid black lines in Figure 3.

*Step 4. Compute the function evaluation (i.e. SF score) for each of the projected simplex vertices to obtain $f(p'_1), f(p'_2), \ldots, f(p'_n)$.*

We can compute the SF score for a particular treatment plan solution $p$ using Equations 1 and 2.

*Step 5. Order the simplex points by their function evaluations such that $f(p'_1) \leq f(p'_2) \leq \ldots \leq f(p'_n)$*

Steps 5-11 are adapted directly from the Nelder-Mead Simplex Search algorithm. Simplex search terminates when convergence criteria are met or the algorithm has reached the maximum number of iterations allowed. We define these termination criteria as (1) the maximum Euclidean distance between simplex vertices must be smaller than 5 Gy and (2) the maximum absolute difference between all plan SF scores must be smaller than 0.05.

*Step 6. Compute the centroid $p_c$ of the $n-1$ best simplex points*

*Step 7. Reflection:*
  a. Compute the reflected point in the decision space as $p_r = p_c + \alpha(p_c - p_n)$. We follow the official MATLAB version and use $\alpha = 1$.
  b. Project the reflected point to the Pareto front to obtain $p'_r$.
  c. Compute the function evaluation $f(p'_r)$.
  d. If $f(p'_1) \leq f(p'_r) < f(p'_{n-1})$, replace the worst simplex point with $p_r$ and start from 5.

*Step 8. Expansion:*
  a. If $f(p'_r) < f(p'_1)$, compute the expansion point in the decision space as $p_e = p_c + \beta(p_c - p_n)$. We follow the official MATLAB version of the Nelder-Mead Simplex Search and use $\beta = 2$.
  b. Project the expansion point to the Pareto front to obtain $p'_e$.
  c. Compute the function evaluation $f(p'_e)$.
  d. If $f(p'_e) < f(p'_r)$, replace the worst simplex point with $p_e$ and start from 5.
  e. Otherwise, replace the worst simplex point with $p_r$ and start from 5.

*Step 9. Outside Contraction:*
  a. If $f(p'_{n-1}) \leq f(p'_r) < f(p'_n)$, compute the outside contraction point in the decision space as $p_{oc} = p_c + \gamma(p_r - p_c)$. We follow the official MATLAB version and use $\gamma = 0.5$.
  b. Project the outside contraction point to the Pareto front to obtain $p'_{oc}$.
  c. Compute the function evaluation $f(p'_{oc})$.
  d. If $f(p'_{oc}) < f(p'_r)$, replace the worst simplex point with $p_{oc}$ and start from 5.
  e. Otherwise, continue to step 11

*Step 10. Inside Contraction:*
  a. If $f(p'_n) \leq f(p'_r)$, compute the inside contraction point in the decision space as $p_{ic} = p_c + \gamma(p_n - p_c)$. We follow the official MATLAB version and use $\gamma = 0.5$.
  b. Project the inside contraction point to the Pareto front to obtain $p'_{ic}$.
  c. Compute the function evaluation $f(p'_{ic})$.
  d. If $f(p'_{ic}) < f(p'_n)$, replace the worst simplex point with $p_{ic}$ and start from 5.
  e. Otherwise, continue to step 11

*Step 11. Shrink:*
  a. Replace every point except the best point (i.e. $p_i$ where $i = 2, \ldots, n$) with $p_i = p_1 + \rho(p_i - p_1)$. We follow the official MATLAB version and use $\rho = 0.5$.
  b. Start from step 5